\documentclass[twocolumn]{aastex63}
\pdfoutput=1
\usepackage[utf8]{inputenc}

\accepted{October 7, 2020}
\submitjournal{\apj}
\newcommand{\msun}{\mbox{$\rm M_\odot$}}

\shorttitle{Mass segregation in galaxy clusters}
\shortauthors{S.W. Kim et al.}

\usepackage{lipsum}
\usepackage{verbatim}
\usepackage{amsmath}

\DeclareUnicodeCharacter{02BC}{'}

\begin{document}

\title{YZiCS: On the Mass Segregation of Galaxies in Clusters}

\author[0000-0002-6787-3020]{Seonwoo Kim}
\affiliation{Department of Astronomy \& Yonsei University Observatory, Yonsei University, Seoul 03722, Republic of Korea}

\author[0000-0002-2873-8598]{Emanuele Contini}
\affiliation{School of Astronomy and Space Science, Nanjing University, Nanjing 210093, Peopleʼs Republic of China}

\author[0000-0001-7229-0033]{Hoseung Choi}
\affiliation{Department of Astronomy \& Yonsei University Observatory, Yonsei University, Seoul 03722, Republic of Korea}

\author[0000-0001-9939-713X]{San Han}
\affiliation{Department of Astronomy \& Yonsei University Observatory, Yonsei University, Seoul 03722, Republic of Korea}

\author[0000-0002-6810-1778]{Jaehyun Lee}
\affiliation{Korea Institute for Advanced Study, 85, Hoegi-ro, Dongdaemun-gu, Seoul 02455, Republic of Korea}

\author[0000-0002-4731-9604]{Sree Oh}
\affiliation{Research School of Astronomy \& Astrophysics, The Australian National University, Canberra, ACT 2611, Australia}
\affiliation{ARC Centre of Excellence for All Sky Astrophysics in 3 Dimensions (ASTRO 3D), Australia}

\author{Xi Kang}
\affiliation{Purple Mountain Observatory, the Partner Group of MPI für Astronomie, 2 West Beijing Road, Nanjing 210008, Peopleʼs Republic of China}

\author[0000-0002-4556-2619]{Sukyoung K. Yi}
\affiliation{Department of Astronomy \& Yonsei University Observatory, Yonsei University, Seoul 03722, Republic of Korea}

\begin{abstract}
Mass segregation, a tendency of more massive galaxies being distributed closer to the cluster center, is naturally expected from dynamical friction, but its presence is still controversial. 
Using deep optical observations of 14 Abell clusters (KYDISC) and a set of hydrodynamic simulations (YZiCS), we find in some cases a hint of mass segregation inside the virial radius. Segregation is visible more clearly when the massive galaxy fraction is used instead of mean stellar mass. The trend is more significant in the simulations than in the observations. To find out the mechanisms working on mass segregation, we look into the evolution of individual clusters simulated. We find that the degree of mass segregation is different for different clusters: the trend is visible only for low-mass clusters. We compare the masses of galaxies and their dark haloes at the time of infall and at the present epoch to quantify the amount of tidal stripping. We then conclude that satellites that get accreted at earlier epochs or galaxies in more massive clusters go through more tidal stripping. These effects in combination result in a correlation between the host halo mass and the degree of stellar mass segregation.

\end{abstract}

\section{introduction}
It is well known that galaxies are not distributed uniformly within (at least) the virialized region, and most of their properties are influenced by their local environments.  Early-type galaxies are usually spheroidal, red, star-formation passive, and occupy the inner regions of clusters, whereas late-type galaxies tend to be disky, blue, star-formation active, and prefer low-density environments such as the field \citep[e.g.,][]{dressler1980}. This behavior is known as the ``morphology-density relation"
\citep{postman1984, gomez2003, kauffmann2004, Balogh2004, weinmann2006, weinmann2010, vonderlinden2010, wetzel2012}.

Morphology, colors, and star formation activity are not the only properties of galaxies for which some type of dichotomy has been found. Especially in the last decade, many authors (e.g., \citealt{presotto2012, Balogh2004, Roberts2015, Contini2015, Joshi2016}, just to quote the latest) have focused attention on stellar mass segregation, with the claim that more massive galaxies tend to be distributed closer to the cluster center, usually for dynamical reasons that are often linked to the correlation between a satellite galaxy and its dark matter halo. If more massive or luminous satellites are associated with more massive haloes, as one would expect, dynamical friction would bring more massive haloes (and thus galaxies) more quickly to the innermost regions of the cluster \citep{chandrasekhar1943}.

In the context of dark matter subhaloes, mass segregation has been discussed extensively by several authors \citep[e.g.,][]{delucia2004,contini2012,vandenbosch2016}. \citet{contini2012} showed that haloes that are more massive at the time of infall to the cluster get closer to the center on shorter timescales due to dynamical friction and suffer more significant stripping. On the other hand, the luminosity and the stellar mass are expected to be more strongly related to the mass of the galaxy at the time of infall than to the properties of the present dark matter haloes \citep{gao2004a,vale2006,wang2006}. Consequently, it is natural to expect galaxies to be segregated according to their stellar mass in clusters.

However, there is not yet a general consensus on these issues; both observationally and theoretically different authors either do find segregation \citep[e.g.,][]{lares2004,mcintosh2005,vandenbosch2008,presotto2012,balogh2014,Roberts2015,Contini2015,Joshi2016,vandenbosch2016,nascimento2017} or else they find weak or no segregation \citep[e.g.,][]{ziparo2013,vulcani2013,vonderlinden2010,kafle2016,joshi2017}.

Among observational and theoretical studies in favor of mass segregation, we quote here the most recent ones. \citet{Roberts2015} used galaxy ``group" catalogs derived from the Sloan Digital Sky Survey Data Release 7 (SDSS DR7) and found clear mass segregation, the strength of which depends on both the galactic stellar mass cut (being higher with the inclusion of lower mass galaxies) and on the ``group" halo mass (decreasing with increasing group halo mass). \citet{Contini2015} took advantage of a semi-analytic model of galaxy formation coupled with a high-resolution N-body simulation explicitly to probe mass segregation in groups and clusters. They found mass segregation in both the stellar mass and the dark matter mass out to the virial radius of the main halo. Using galaxy analogues related to N-body simulations \citet{Joshi2016} found a behavior similar to the results shown by \citet{Contini2015} up to 0.5 $R_{\rm vir}$ of the host halo, but with a weaker trend with increasing radial distance. Clear evidence of mass segregation in the dark matter mass has been found also by \citet{vandenbosch2016}. They ran three cosmological N-body simulations and focused on the segregation of 12 properties of subhaloes that depend upon the orbital energy and distance from the halo center, and they found that subhaloes do not show any segregation when their present-day masses are considered. However, when their mass at infall is considered, the segregation in mass is evident. More massive subhaloes fall faster to the center because of stronger dynamical friction, and lose a larger fraction of mass via tidal stripping. As discussed above, these objects are expected to host massive galaxies.

Other authors have found different results. \cite{ziparo2013} studied 52 X-ray selected galaxy groups at $0 < z < 1.6$ and found no clear evidence of mass segregation. Similar results have been found by \cite{vonderlinden2010} at $z < 0.1$ in a sample of SDSS galaxy clusters, using different stellar mass cuts in four redshift ranges. Indirectly, \cite{vulcani2013} worked on the galaxy stellar mass function at intermediate redshift, using bins of distance from the cluster center in different environments, and they found no sign of mass segregation. More recently, \cite{kafle2016} presented a full comparison between the observed Galaxy And Mass Assembly (GAMA) data, a semi-analytic model ({\small GALFORM}), and the EAGLE hydrodynamic simulation data and found negligible mass segregation in galaxy group environments. Similar conclusions have been obtained by \cite{joshi2017}, who found weak mass segregation in the inner regions of groups and clusters.

In this work, we address this topic by taking advantage of a set of zoom-in hydrodynamic simulations and a set of observed galaxy clusters in the local universe. Our goal is to analyze the distributions of galaxies in clusters within 1 $R_{\rm vir}$, by looking at both the stellar and dark matter masses of galaxies and their associated haloes.

This manuscript is structured as follows. In Section 2, we describe the main features of our simulations and the observational data. In Section 3, we present our results, which are discussed in Section 4. We summarize the main conclusions of our analysis in Section 5.

\section{Data}

In this work, we use both an observational and a theoretical approach. We take advantage of a set of observational data of galaxy clusters, named the KASI-Yonsei Deep Imaging Survey of Clusters \citep[KYDISC,][]{Oh2018}, and a set of hydrodynamic zoom-in cluster simulations, named the Yonsei Zoom-in Cluster Simulations \citep[YZiCS,][]{Choi2017}. Their main features are presented below.

\subsection{Observational Data}

KYDISC was performed with the Inamori Magellan Areal Camera and Spectrograph \citep[IMACS,][]{Dressler2006} on the 6.5-meter Magellan Baade telescope at Las Campanas Observatory (LCO) and the MegaCam on the 3.6-meter Canada-France-Hawaii Telescope to target 14 Abell clusters at $0.015 < z < 0.144$.

Then, follow-up spectroscopy was done using the Magellan IMACS, the Wide-field Reimaging CCD Camera (WFCCD) on the du Pont 2.5 m telescope at LCO, and Hydra on the WIYN 3.5 m telescope.
The redshift information of galaxies was attained from the literature (141 galaxies from the NASA/IPAC Extragalactic Database; 619 from the SDSS; and 380 from the Hectospec Cluster Survey; 5 from the 6dF survey) and derived from the observations (112 galaxies from the IMACS; 9 from the du Pont; and 143 from the WIYN).
This results in magnitudes, redshifts, morphologies, bulge-to-total ratios, and local densities for the 1409 cluster galaxies brighter than $-19.8$ in the r-band. The spectroscopic completeness is 0.8 if Abell 1278 is excluded. To derive the stellar mass of cluster member galaxies, we use the equation from \citet{Bell2003}.
Since the formula gives $M_{\rm\ast}$ based on a ``diet" Salpeter initial mass function (IMF), we have multiplied 0.7 to $M_{\rm\ast}$ in order to convert it to a normal Salpeter IMF. We consider the point at which the low mass end of the stellar mass function begins to decline as the mass threshold for our analysis, which is $M_{\rm\ast}$ = $10^{10.3}\ M_{\rm\odot}$ in this case. We use the position of the brightest cluster galaxies (BCGs) as the cluster centers, and we remove the BCGs from our analysis. For both sets of data we used $R_{\rm 200}$ as the virial radius. For more detailed information about the KYDISC data, we refer the reader to \citet{Oh2018}.

\subsection{Hydrodynamic Simulations}

YZiCS is a set of cosmological zoom-in simulations performed with the adaptive mesh refinement code RAMSES \citep{Teyssier2002}. Dark matter only simulation was first run within a cubic volume 200 $\rm h^{-1}$ Mpc on each side. In this parent volume, a total number of 16 dense regions have been selected. The selected volume out to 3 $R_{\rm vir}$ were backtraced until the initial condition, and the hydro zoom-in simulation was conducted for the spherical volume that contains all the particles. The simulation was based on the same baryon physics recipe that was used in the Horizon-AGN simulation \citep{Dubois2012}, including feedback from both active galactic nuclei (AGNs) and supernovae (SNe). Throughout the simulations, the WMAP7 cosmology from \citet{Komatsu2011} was assumed: ${\Omega}_{\rm m}$ = 0.272, ${\Omega}_{\rm\lambda}$ = 0.728, $H_{\rm 0}$ = 70.4 km $\rm s^{-1}\ Mpc^{-1}$, ${\sigma}_{\rm 8}$ = 0.809, and n = 0.963. Objects with more than 200 stellar particles were classified as galaxies, which roughly corresponds to $10^{8}\ M_{\rm\odot}$. We identified a total of 6656 galaxies above the mass threshold $M_{\rm\ast}$ = $10^{9}\ M_{\rm \odot}$ by using the AdaptaHOP halo finder \citep{Aubert2004}. The stellar mass threshold corresponds roughly to the point at which the stellar mass function begins to decline at the low mass end, and we consider it to be the value of the stellar mass to which our galaxy sample is reasonably complete. For more details concerning the properties and features of the simulation, we refer the reader to \citet{Choi2017}.

We have created two catalogs with YZiCS: one is drawn only from the simulations, and the other is adjusted for comparison with the observations. In the first catalog, the distance of a galaxy from the cluster center is measured by projecting into two dimensions (2D) in random directions. To make fair comparisons with the observations, we defined the BCGs as the cluster centers and made 100 random projections into 2D. This results in a total of 665,600 projected properties observed in the simulations. In order to look back in time and inspect various environmental effects, in the second catalog we used the galaxy merger tree constructed by \citet{Lee2018}, which was created using the algorithm given by \citet{jung2014}. To match galaxies to the haloes in which they reside, we made use of the linking length motivated by \citet{Behroozi2013}, where
\begin{equation}
d(h,p) = \left(\frac{{|x_{\rm h}-x_{\rm g}|}^{2}}{{r^{2}_{\rm{dyn,vir}}}}+\frac{{|v_{\rm h}-v_{\rm g}|}^{2}}{\sigma_{\rm v}^{2}}\right)^{1/2}
\end{equation}
is the distance in the 6D phase-space of positions and velocities. The quantities $x_{\rm h}$ and $v_{\rm h}$ are the halo position and velocity, $r_{\rm{dyn,vir}}$ is the virial radius of the halo, $\sigma_{\rm v}$ is the velocity dispersion of the halo, and $x_{\rm g}$ and $v_{\rm g}$ are the galaxy position and velocity. To derive the velocity dispersion of the galaxies included within the linking length, we used the mass-velocity dispersion relation from \citet{munari2013}:
\begin{equation}
\frac{\sigma_{\rm 1D}}{\rm km \ \rm s^{-1}} = A_{\rm 1D} \left[ \frac{h(z)M_{\rm 200}}{10^{15}M_{\rm\odot}} \right]^{\alpha}
\end{equation}
with $A_{\rm 1D}$ = 1244 km\ $\rm s^{-1}$ and $\alpha$ = 1/3.

We sorted the list of galaxies and chose the subhalo with the shortest linking length as the host for each galaxy. We performed this matching process both for galaxies at redshift zero and at their time of infall; i.e., when they crossed 1.5 $R_{\rm vir}$ for the last time. We set the distance at infall to be 1.5 $R_{\rm vir}$ to prevent cluster tides from affecting the galaxies and subhaloes before infall. Although galaxies may cross this baseline many times, we used this criterion in order to remove back-splashed galaxies from `accreted' galaxy samples. This is because galaxies that move out of the cluster are not defined as a part of the accreted portion of the cluster. To make sure that the subhaloes matched at redshift zero and at the time of infall are from the same subhalo tree, we checked manually for sudden changes in mass and position. Unlike the first catalog, in this case the distances of galaxies from their cluster centers are in 3D, and we considered the location of the main haloes with the maximum number of dark matter particles to be the cluster center. Throughout this paper, we removed the BCGs from our analysis. The resulting information of the KYDISC and the YZiCS clusters
is presented in Tables 1 and 2. In Section 3.1 below, we make use of the first catalog, while in Section 3.2, we use the second one described above.

\section{results}
In this section we analyze our observational (KYDISC) and simulation (YZiCS) data to see whether or not they show any mass segregation trend. For fair comparisons, we used the same mass threshold ($10^{10.3}\ M_{\rm \odot}$) to them.

\begin{deluxetable}{lccc}[t!]
\tablecaption{KYDISC Cluster Characteristics}
\tablecolumns{4}
\tablenum{1}
\tablewidth{0pt}
\tablehead{
\colhead{Name} &
\colhead{$M_{\rm vir}$ \tablenotemark{a}} &
\colhead{$R_{\rm vir}$\tablenotemark{b}} &
\colhead{$N_{\rm sat}$\tablenotemark{c}} \\
\colhead{} & \colhead{($10^{14}\ M_{\rm\odot}$)} &
\colhead{(Mpc)} & \colhead{$\geq 10^{10.3}\ M_{\odot}$}
}
\startdata
Abell 116 & 2.25 & 1.26 & 37 \\
Abell 646 & 5.33 & 1.64 & 93 \\
Abell 655 & 6.43 & 1.75 & 166 \\
Abell 667 & 5.12 & 1.61 & 113 \\
Abell 690 & 3.42 & 1.44 & 76 \\
Abell 1126 & 4.99 & 1.63 & 61 \\
Abell 1139 & 1.25 & 1.04 & 40 \\
Abell 1146 & 10.83 & 2.07 & 115 \\
Abell 1278 & 9.54 & 1.99 & 41 \\
Abell 2061 & 11.40 & 2.15 & 221 \\
Abell 2249 & 12.39 & 2.21 & 226 \\
Abell 2589 & 10.40 & 2.11 & 61 \\
Abell 3574 & 1.92 & 1.21 & 8 \\
Abell 3659 & 2.04 & 1.21 & 20 \\
\enddata
\tablenotetext{a}{Virial mass of the clusters.}
\tablenotetext{b}{Virial radius of the clusters.}
\tablenotetext{c}{The number of satellites within 1 $R_{\rm vir}$ of each of the cluster with the mass cut $10^{10.3}\ M_{\rm\odot}$.}
\end{deluxetable}

\begin{deluxetable}{lcccc}[t!]
\tablecaption{YZiCS Cluster Characteristics}
\tablecolumns{6}
\tablenum{2}
\tablewidth{0pt}
\tablehead{
\colhead{Name} &
\colhead{$M_{\rm vir}$ \tablenotemark{a}} &
\colhead{$R_{\rm vir}$\tablenotemark{b}} &
\colhead{$N_{\rm sat}$\tablenotemark{c}} &
\colhead{$N_{\rm sat}$\tablenotemark{d}} \\
\colhead{} & \colhead{($10^{14}\ M_{\rm\odot}$)} &
\colhead{(Mpc)} & \colhead{$\geq 10^{10.3}\ M_{\odot}$} & \colhead{$\geq 10^{9}\ M_{\odot}$}
}
\startdata
C1 & 9.02 & 1.99 & 106 & 489 \\
C2 & 5.12 & 1.64 & 45 & 245 \\
C3 & 2.06 & 1.21 & 27 & 149 \\
C4 & 1.99 & 1.20 & 17 & 110 \\
C5 & 1.98 & 1.20 & 21 & 143 \\
C6 & 1.93 & 1.19 & 19 & 87 \\
C7 & 1.78 & 1.16 & 24 & 159 \\
C8 & 1.53 & 1.10 & 14 & 93 \\
C9 & 1.36 & 1.06 & 17 & 93 \\
C10 & 1.30 & 1.04 & 14 & 91 \\
C11 & 0.68 & 0.84 & 6 & 65 \\
C12 & 0.59 & 0.80 & 8 & 34 \\
C13 & 0.55 & 0.78 & 3 & 40 \\
C14 & 0.50 & 0.76 & 7 & 34 \\
C15 & 0.46 & 0.74 & 3 & 32 \\
C16 & 2.51 & 1.30 & 34 & 164 \\
\enddata
\tablenotetext{a}{Virial mass of the clusters.}
\tablenotetext{b}{Virial radius of the clusters.}
\tablenotetext{c}{The number of satellites within 1 $R_{\rm vir}$ of each of the cluster with the mass cut $10^{10.3}\ M_{\rm\odot}$.}
\tablenotetext{d}{The number of satellites within 1 $R_{\rm vir}$ of each of the cluster with the mass cut $10^{9}\ M_{\rm\odot}$.}
\end{deluxetable}

\subsection{Galaxy Mass Distribution in 2D Projection}

\begin{figure}[t!]
\includegraphics[width=8.5cm]{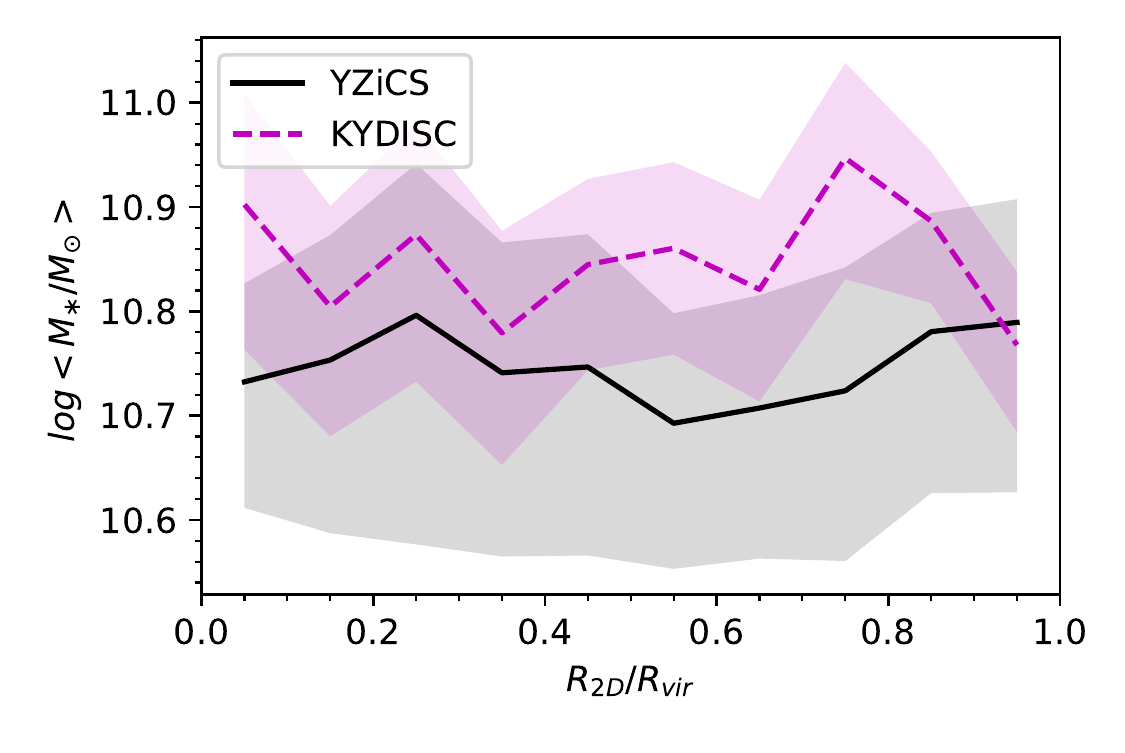}
\centering
\caption{The mean stellar mass of the KYDISC and YZiCS galaxies as a function of 2D clustocentric distance. See the text for details. The magenta and gray shades represent the 1${\sigma}$ scatter around the mean values. The mass threshold is $10^{10.3}\ M_{\rm \odot}$ for both the KYDISC and the YZiCS data.}
\end{figure}

Figure 1 shows the mean stellar mass in each radial shell from KYDISC (magenta line) and YZiCS (black line). The region $R_{\rm 2D} \leq 1 \ R_{\rm vir}$ in each cluster was divided into 10 distance bins, which results in 10 concentric shells. The two lines represent the mean values of 16 YZiCS and 14 KYDISC clusters, while the shaded regions represent 1$\sigma$ scatter around the mean. Neither of them clearly shows mass segregation. The lack of mass segregation we find here may be due to several factors. Firstly, to enhance the completeness of our samples we do not include low-mass galaxies which are suspected to reveal the mass segregation trend better \citep{Roberts2015}, and secondly, some of our clusters lie in the highest range of cluster masses compared to those presented in the literature, where mass segregation effect is found to be either weak or null (see Section 3.2). We discuss these two effects in more detail in Section 3.2.

Another complication in the analysis shown in Figure 1 is that it does not take into account the fact that the distribution of galaxy masses depends on the cluster mass. Mixing different galaxy mass distributions of clusters of different masses would hide a trend even though there is any. Besides, Figure 1, which is the format that has been widely used in the literature, shows the mean galaxy mass in each radial bin. In this case, even if more of the massive galaxies are distributed in the inner volume in a more massive cluster, the ``mean'' stellar mass might still be smaller, since there are many more of less massive galaxies in the inner regions.

In order to avoid such effects of cluster masses, we have decided to use a fixed ``fraction'' of more massive galaxies within the sample of each cluster instead of mean masses. For example, if we use 50\% as our cut, we count the top 50\% massive galaxies in each cluster and present their fraction as a function of radial distance. We perform this for all clusters with varying cluster masses. Then, we have the mean radial distribution of the 50\% most massive galaxies in each cluster for all clusters. For reference, the typical mass criterion for the top 50\% is higher than $10^{11}\ \msun$. In the case of KYDISC (or YZiCS), we have 14 (16) data points as a result in each radial distance bin, and we derive their mean and error through the bootstrapping method. Figure 2 shows the result. The mass segregation trend has become clearer at least in the case of YZiCS. KYDISC shows a less clear trend, and we discuss this in Section 4.

\begin{figure}[t!]
\includegraphics[width=8.5cm]{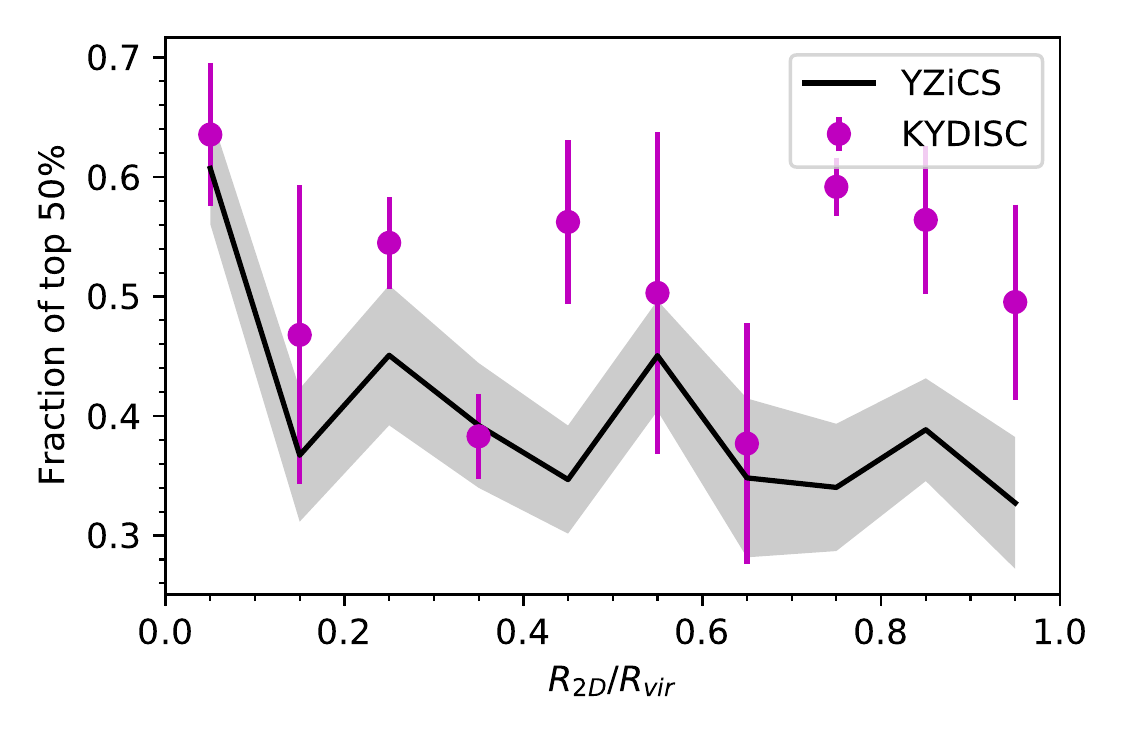}
\centering
\caption{The fraction of massive galaxies (the top 50\% of the stellar mass distribution) relative to the total number of galaxies more massive than $10^{10.3}\ M_{\rm\odot}$ as a function of clustocentric distance for the whole sample of KYDISC and YZiCS clusters. The error bars and the shaded region indicate the 1$\sigma$ scatter.}
\end{figure}

\begin{figure}[t!]
\includegraphics[width=8.5cm]{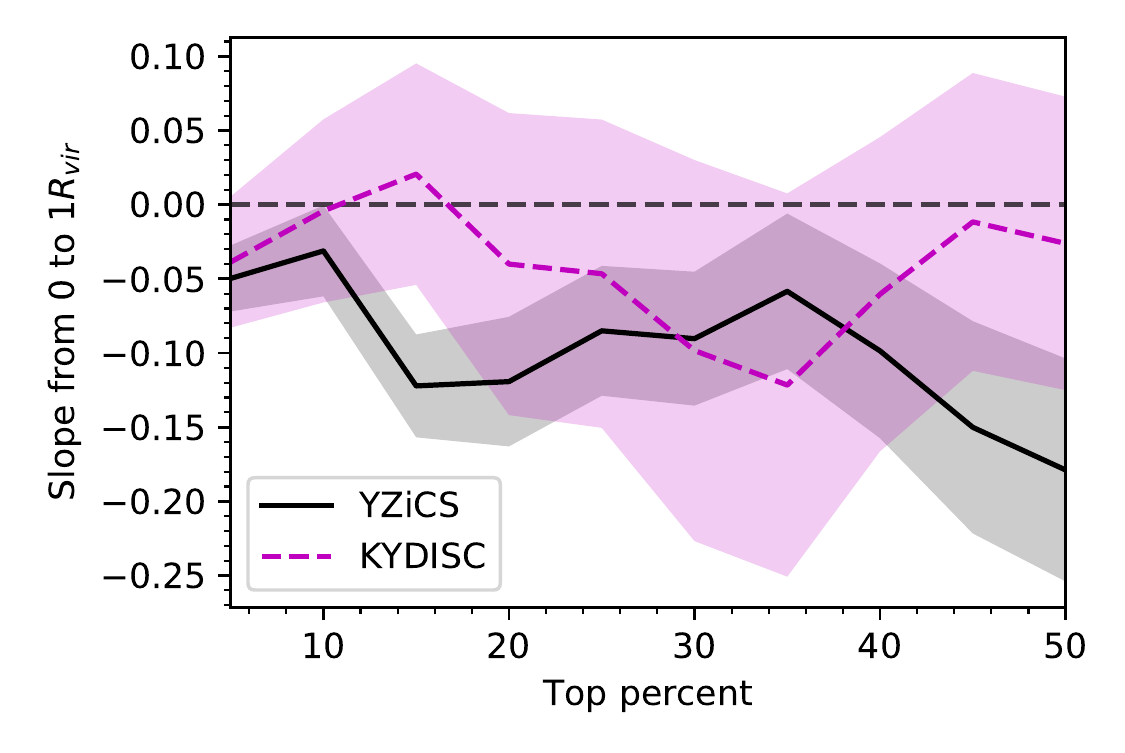}
\centering
\caption{Slopes of the linear fits to the data points shown in Figure 2 based on the use of different fractional cuts. Figure 2 shows the case of 50\%. The shaded regions represent the 1$\sigma$ scatter of the distribution. A negative slope indicates the presence of mass segregation. The YZiCS data consistently suggest mass segregation regardless of the choice of the fractional cut. The KYDISC data, on the other hand, suggests a hint of mass segregation for the cut of 20\% or more, because the uncertainty band is consistently below the zero slope.}
\end{figure}

Our choice of the fractional cut of 50\% in mass is arbitrary, but the analysis is stable against the choice as long as a reasonable value is used. We fitted the mass segregation slope of Figure 2 in the range 0 $\leq R_{\rm 2D}/R_{\rm rvir} \leq$ 1 to the linear relation $\alpha$x+$\beta$ using the ${\chi}^{2}$--minimization method, with different percentiles for the high galaxy mass cut, and the slopes `$\alpha$' for different massive galaxy percentiles are shown in Figure 3. Having a negative slope means that mass segregation does exist. Mass segregation appears robust for the YZiCS data regardless of the choice of the fractional mass criterion. It is less clear for the KYDISC data and marginally visible for the range of $\gtrsim 20\%$, because the uncertainty band tends to locate below zero. Considering the limited number of sample galaxies above the mass cut of $10^{10.3} \ \msun$ (see Tables 1 and 2), a fractional cut of $\gtrsim 20\%$ is recommended.

\subsection{Halo Mass Dependence of Mass Segregation}

\begin{figure}[t!]
\includegraphics[width=8.5cm]{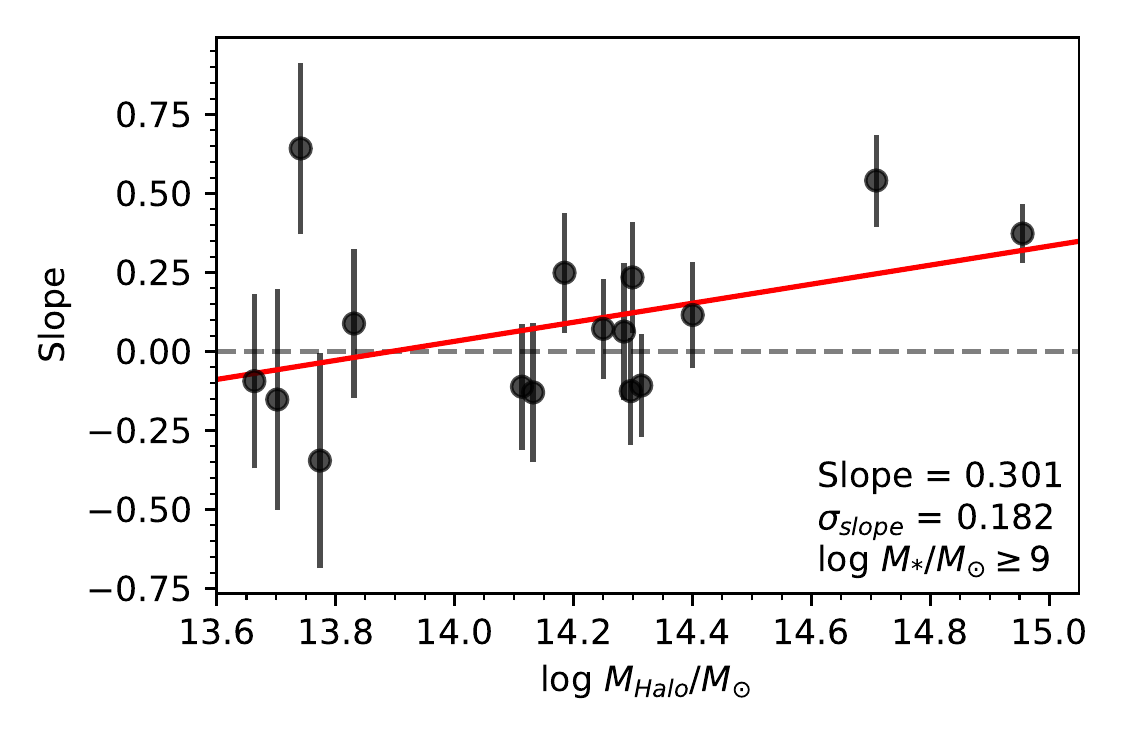}
\caption{Mass segregation slope of the stellar mass and radial distance relationship as a function of halo mass; the dots represent the slopes and the error bars the associated 1$\sigma$ scatter for each of our 16 YZiCS clusters; the red line is the result of an ${\chi}^{2}$--fit, and its slope and error are shown in the legend. The slopes in this figure are different from that of Figure 3 because all of the galaxy mass over $10^{9}\ M_{\rm\odot}$ were used without clustocentric distance binning. The cluster mass ranges 0.46 $\times 10^{14}\ M_{\odot}$ -- 9.02 $\times 10^{14}\ M_{\odot}$.}
\end{figure}

The mass segregation trend was not visible in the mean stellar mass vs. clustocentric distance diagram (Figure 1). However, mass segregation became visible when we properly normalize the mass functions of galaxies in different clusters based on mass ranking and used the massive galaxy fraction instead of the mean stellar mass (Figure 2). 
To give a physical explanation of why mass segregation happens, we use YZiCS to go back in time and monitor the related quantities to the galaxies.
We have decided to use galaxy mass vs. clustocentric distance on individual clusters in this analysis because too few galaxies are in some YZiCS clusters to use the fractional approach. 

Figure 4 shows the linear-fitted slope of the stellar mass as a function of clustocentric distance for {\em individual} clusters with different halo masses from YZiCS. The fit was made on the unbinned data.
Because we do not compare simulations against observations any more, we use for better statistics a lower mass cut for the galaxies, that is, $M_* \geq 10^{9}\ M_{\rm\odot}$, which is roughly the completeness limit of the YZiCS simulation.
The haloes of smaller masses show more negative slopes, which means that less massive haloes show more mass segregation. 
This is tightly linked with the dynamical friction timescale. 
For a given cluster halo mass, a more massive galaxy reaches the cluster center more quickly.
If we simplistically assume that galaxies with different masses arrive at a cluster at the same time, therefore, radial mass segregation is naturally expected after some time. 
This may be what happened to low-mass clusters.

It becomes more complicated in massive clusters, however. According to the same dynamical friction formula, a galaxy of a given mass takes longer to reach the center of a more massive cluster. During this longer period, the galaxy is bound to lose more mass due to stellar stripping \citep{Smith2016}.
To consider this cluster mass effect in our analysis, we divide the YZiCS cluster sample into two groups: the most massive two clusters (C1 and C2 in Table 2) in one group and the rest in the other, simply because the mass distribution of the clusters shows a large gap between them. 
Figure 5 shows the mean galaxy mass as a function of clustocentric distance for the two groups. Two mean galaxy masses are shown: those at the present epoch marked by the tips of the arrows, and those at the time of infall (when they cross 1.5 $R_{\rm vir}$ for the last time) marked by the bases of the arrows. The arrows show the difference between the galaxy mass at the infall and the current epoch. 
The first to note is the mean mass of the galaxies in the central bin of the less massive clusters which is significantly higher than any other points in the diagram. We interpret this as a direct result of dynamical friction.

The infall masses of the galaxies in the central bins of the more massive clusters (three leftmost magenta points), however, are not particularly higher than the values in other bins. This is because they arrived at their clusters relatively earlier, as demonstrated in Figure 6. This figure shows the kernel density estimation of the redshift at infall of galaxies in the two halo mass groups. We consider the galaxies in each cluster as samples from the same population. In reality, clusters are different from each other due to the difference in the detailed history as we are demonstrating in this study. In this sense, our assumption that all the cluster galaxies are from the same parent sample is overly optimistic, and our errors here are upper limits. 
Having that in mind, the difference in $z_{\rm inf}$ of the two subsamples appear marginally distinct. Although the scatter do overlap between the two subsamples, more massive clusters have consistently higher values of $z_{\rm inf}$ at $R \lesssim 0.4 \ R_{\rm vir}$ (Figure 6).
 
Because the galaxies in the inner volume of more massive clusters spend a longer time inside their cluster than those of the less massive clusters, they lose more stellar mass (e.g., magenta arrows at $< 0.3 \ R_{\rm vir}$ in Figure 5). In conclusion, the effect of mass segregation is at work here in the more massive clusters as well, but it is invisible because of the competing effect of arrival time being earlier for the galaxies in the central region. 

\begin{figure}[t!]
\includegraphics[width=8.5cm]{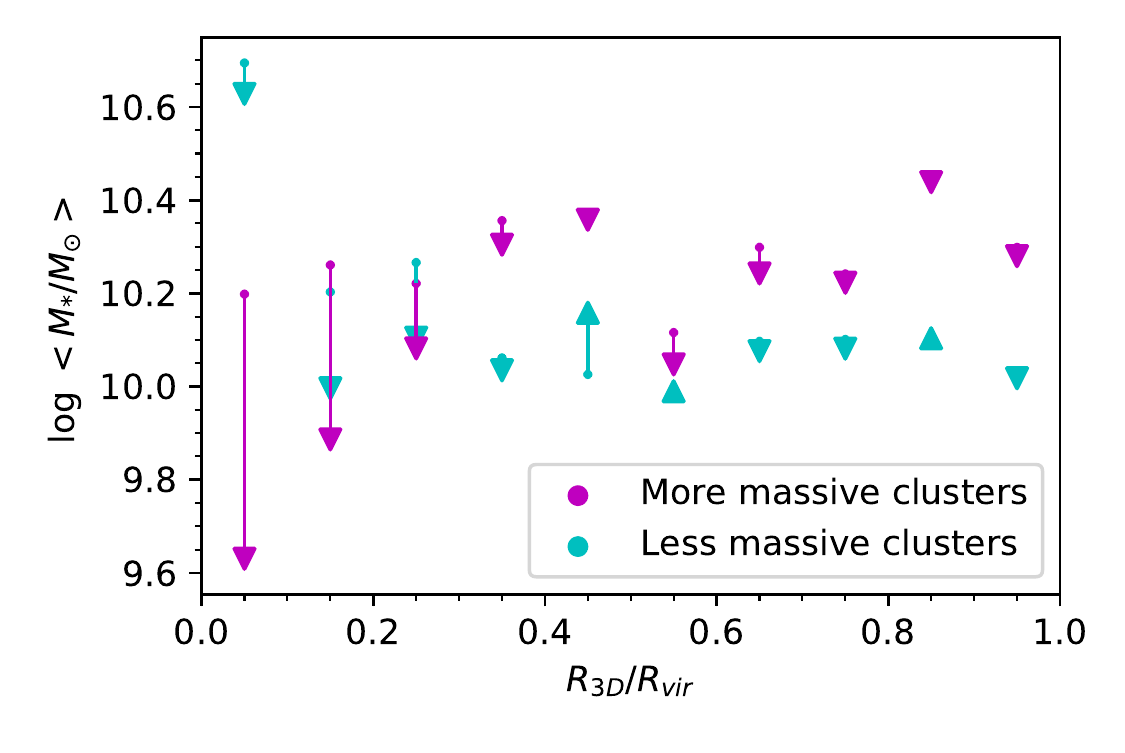}
\caption{Mean stellar mass of each clustocentric shells at the time of infall (bases of the arrows) and present epoch (tips of the arrows), for the 14 less massive clusters and 2 more massive clusters. The inner volume of both subsamples lost more stellar mass than the outskirts, and the galaxies in the inner volume of the more massive clusters lost more mass than the less massive clusters.}
\end{figure}
\begin{figure}[t!]
\includegraphics[width=8.5cm]{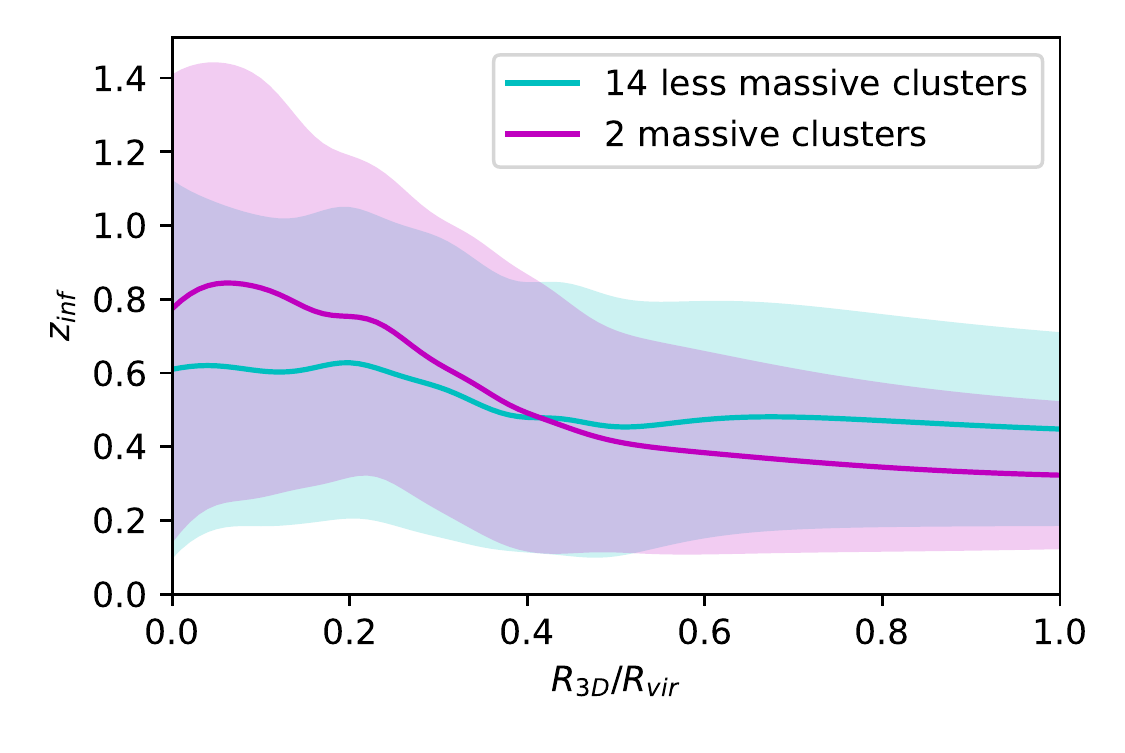}
\caption{Kernel density estimation of redshift at infall ($z_{\rm inf}$) of galaxies in each of the two halo mass groups. The bandwidth is determined by using the Silverman's rule of thumb \citep{Silverman1986}. Both of the subsamples show a decrease of redshift at infall as a function of clustocentric distance and the more massive clusters show higher redshift at infall in general.}
\end{figure}

\begin{figure}[t]
\includegraphics[width=8.5cm]{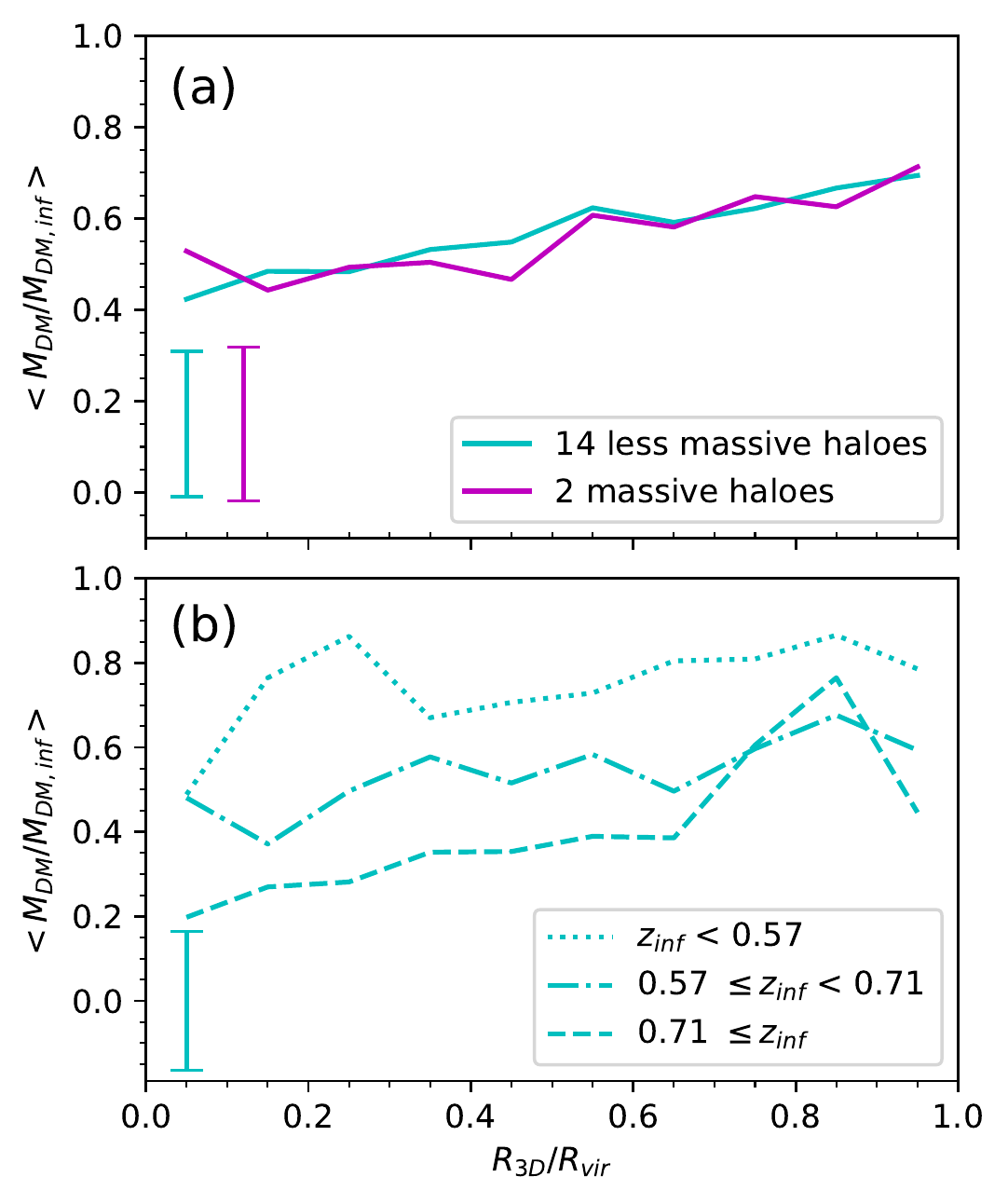}
\centering
\caption{Subhalo mass at the present epoch ($M_{\rm DM}$) divided by subhalo mass at infall ($M_{\rm DM,inf}$) from YZiCS as a function of clustocentric distance; (a) for groups of different halo mass clusters, (b) for different bins of redshift at infall. The division of the sample has been made to achieve the same subsample size for statistical conveniences. The typical scatter, shown as vertical bars in each panel, is (a) 0.32 for the 14 clusters; 0.34 for the other, (b) 0.33.}
\end{figure}

The dynamical evolution of galaxies inside a cluster, including dynamical friction, is in fact more dominantly determined by the halo properties of galaxies rather than their baryonic properties. Therefore, what has been presented and discussed above should be visible in the dark matter halo properties of galaxies as well. Figure 7 shows the halo properties of galaxies inside clusters. Panel (a) shows the ratio between the present-day halo mass and the halo mass at the time of infall as a function of clustocentric distance. As expected, the galaxies in the inner regions have less of their dark matter haloes remaining due to tidal stripping for a longer time inside the cluster. 
Panel (b) indeed shows that the subhaloes that fell in the cluster earlier lose more mass. This is consistent with the earlier studies \citep{rhee2017, Han2018}.

\section{discussion}
Galaxy evolution can be explained either by a `\emph{nature}' or a `\emph{nurture}' scenario. In the context of stellar mass segregation in clusters, a \emph{nature} scenario links the galaxy to its mass at the time it was born, and a \emph{nurture} scenario links mass segregation to what happened from the time of accretion to the present time. Dynamical friction steals the orbital energy of massive galaxies thereby making them spiral inward to the cluster center. This happens because objects orbiting in the cluster slower than the satellite pull the satellites backward, thereby causing a sort of `friction' that results in the loss of the satellite's orbital energy.
Using the dynamical friction formula from \citet{chandrasekhar1943}, we find
\begin{equation}
t_{\rm df} \propto \frac{m_{\rm host}}{m_{\rm sat}}
\end{equation}
where $m_{\rm host}$ is the mass of the host halo, and $m_{\rm sat}$ is the mass of the satellite galaxy it contains.
For a given galaxy mass the dynamical friction timescale is therefore shorter for less massive haloes. 

Dynamical friction, however, may not always result in visible mass segregation mainly because galaxy mass is not conserved in some cases. Tidal stripping associated with dynamical friction is effective both for galaxies and their dark haloes.
But the latter are more prone to it because galaxies are in some sense ``protected'' by their host haloes. 
It has been suggested that the tidal stripping of dark haloes significantly precedes that of baryonic galaxies \citep{Smith2016}. 
But eventually, the galaxies that spent a large amount of time inside their cluster lose some mass.
For stellar mass segregation to be significant, the effect of dynamical friction on the subhalo mass must be larger than the effect of tidal stripping on the galaxy mass.
If this is not the case, mass segregation would not be clearly visible, even though it actually happened.

\begin{figure*}[t!]
\includegraphics[width=18cm]{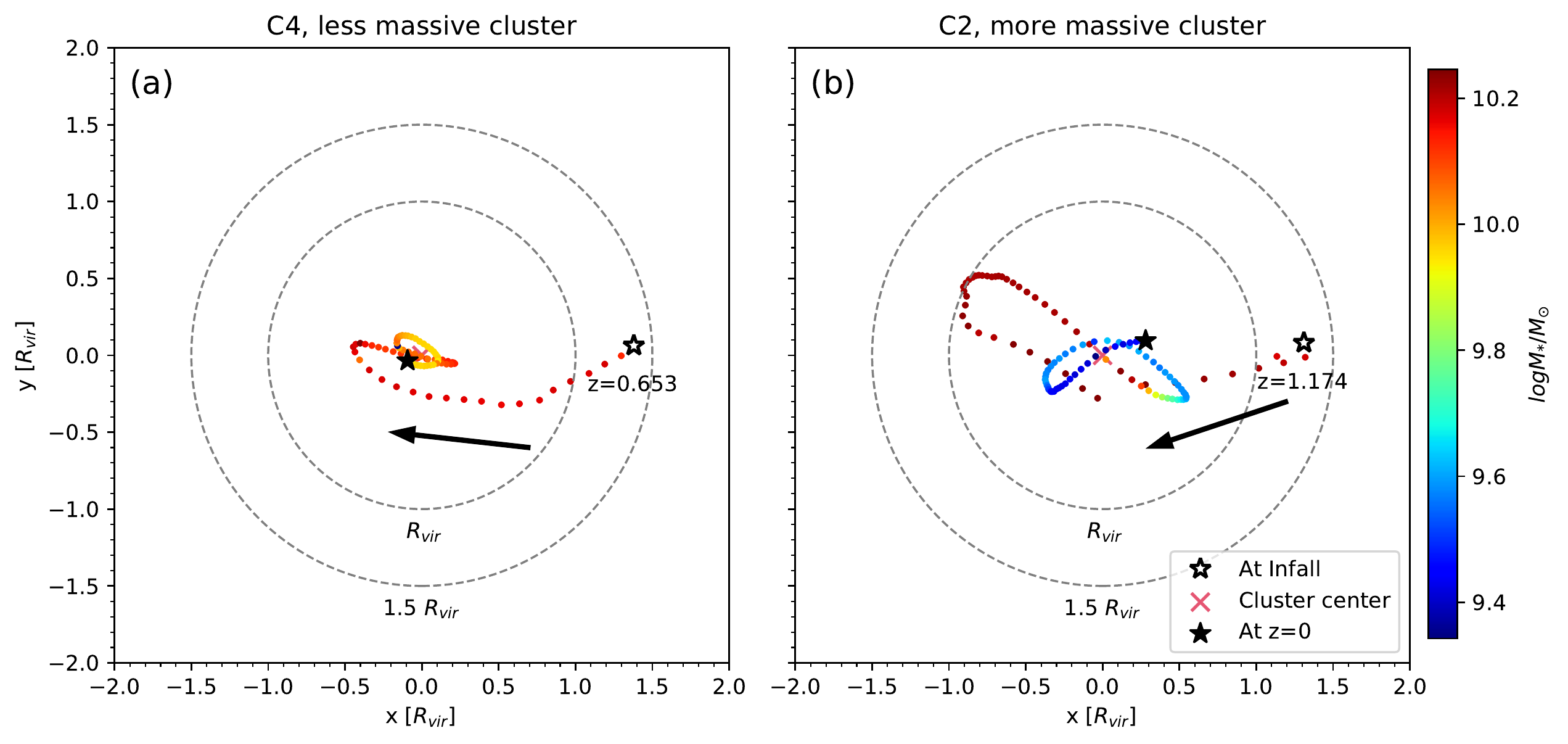}
\centering
\caption{An example of the galaxies in the (a) less massive cluster C4 ($1.99 \times 10^{14}\ M_{\rm\odot}$) and (b) more massive cluster C2 ($5.12 \times 10^{14}\ M_{\rm\odot}$). The empty stars denote the position of the galaxies at the time of accretion, the filled stars indicate the same at the current epoch, and the color coded dots are the same as the epoch in between the former two. The colorbar indicates the stellar mass at each epoch of the two galaxies. To efficiently demonstrate the orbit and the mass changes of the two galaxies, each of the galaxies was projected onto a plane in time sequence, respectively. The normal vector of the projected plane was derived by using the position vector of the galaxy position at the current epoch and at the time of infall, with their vector origin as the cluster center at the present epoch respectively. Dynamical friction works on both of the clusters, but stellar mass segregation is seen only for (a). (b) shows no mass segregation because of tidal stripping: satellites in more massive clusters lose a greater amount of dark matter.}
\end{figure*}

We showed in Figure 4 a robust trend in the segregation slope according to halo mass. To see and quantify what makes and diminishes segregation, we divided the 16 clusters into two groups using $10^{14.6}\ M_{\rm \odot}$ as the halo mass cut. We found that the stellar mass loss due to tidal stripping is greater for the more massive clusters and in the inner regions of both halo groups. This is due to the longer time since infall for galaxies found in the inner regions and for the more massive clusters, as shown in Figure 6. 
This is schematically illustrated in Figure 8 by using the YZiCS data. Panel (a) shows a trajectory of a galaxy that falls into a less massive cluster at $z=0.653$. Its stellar mass, color coded following the key on the right, shows a small decrease through the orbital motion due to tidal stripping. Panel (b) shows another galaxy in a more massive cluster. It was accreted into the cluster much earlier at $z=1.174$ with a similarly large value of mass. But through a long period during its orbital motion, it loses a much larger fraction of mass.
Also, according to Figures 5 and 6, we can infer that less massive galaxies were accreted onto clusters at an earlier epoch in more massive clusters.
These effects combined make the mass segregation effect invisible in massive clusters.

The radial mass segregation of galaxies is linked to the fate of the dark matter haloes that contain them, and to the way they approach the center of their cluster halo. Therefore we use Figure 7, where we show the ratio of the dark matter mass remaining in a subhalo at the present epoch compared to the dark matter mass at infall, when they cross 1.5 $R_{\rm vir}$. 
Haloes that have earlier infall times or are closer to the cluster center go through more mass loss via tidal stripping. Therefore, satellites that are more massive at infall get to the cluster center faster and lose their mass more quickly. Also, the satellites that have early infall time and have managed to exist at the present epoch tend to be less massive at the time of infall. They go through a great amount of mass loss due to tidal stripping. On the other hand, the more massive satellites that have higher redshift at the time of infall have a high possibility of being already merged and thus not appearing in the figures. These two effects together may weaken the stellar mass segregation. We minimized this effect in Figure 2 by looking at the ratio of the more massive galaxies in concentric shells in each of the clusters. This greatly increased the segregation trend for YZiCS, while KYDISC showed a slight amount of mass segregation. The segregation trend of KYDISC may show less mass segregation than it actually has due to the following reasons: projecting clustocentric distances from 3D to 2D lessens the segregation trend (Figures 1 and 5); a relatively high mass cut is used ($10^{10.3}\ M_{\rm\odot}$); the average halo mass range is larger than that of YZiCS; some clusters might not be fully relaxed; and finally KYDISC must have a fair level of foreground and background galaxy contamination.

Previous work on this topic can be understood in this same context. For example, it might be that \citet{vonderlinden2010} found no stellar mass segregation because they have a large galaxy mass cut ($10^{9.6}\ M_{\rm\odot}$). \citet{ziparo2013} find weak stellar mass segregation in the inner regions of the clusters. Their segregation trend is weak when $10^{10.3}\ M_{\rm\odot}$ is used as the mass cut, but the trend becomes much stronger when $10^{9}\ M_{\rm\odot}$ is used as the mass cut. \citet{kafle2016} states that they find a lack of stellar mass segregation because the dynamical friction timescale is longer than the crossing timescale. However, their lack of the trend may instead be the result of using a higher redshift range. 
\citet{vulcani2013} found no stellar mass segregation, possibly because they used a high mass cut of $10^{10.5}\ M_{\rm\odot}$. \citet{vandenbosch2016} find dark matter mass segregation for the accretion mass and peak mass, but this trend disappears when they reach redshift zero because of tidal disruption and the fact that some subhaloes of larger mass get accreted later.
We assume that the absence of mass segregation in their study is due to a different use of the mass cut, in the sense that they used a ratio of the haloes and subhaloes instead of considering the mass functions of each of the haloes.

\section{summary}

In this paper, we have investigated the existence of mass segregation of galaxies in clusters using both observations (KYDISC, cluster mass range $1.92 \times 10^{14}\ M_{\rm\odot}$ -- $1.24 \times 10^{15}\ M_{\rm\odot}$) and a set of hydrodynamic simulations (YZiCS, cluster mass range $0.46 \times 10^{14}\ M_{\rm\odot}$ -- $9.02 \times 10^{14}\ M_{\rm\odot}$). We found mass segregation depending on the cluster properties, and our main results are the following.
\begin{itemize}
    \item When {\em mean stellar mass} is used for the investigation of mass segregation, segregation effect is invisible for both the KYDISC and YZiCS data (Figure 1). 
    \item The segregation effect became visible when {\em massive galaxy fraction} based on the mass rank in each cluster is used instead. This is a way of normalizing the galaxy mass function in terms of cluster mass (Figures 2 and 3).
    \item The stellar mass segregation trend obtained by using all the main haloes of different masses together smears out the trend because individual haloes have trends of different degrees. We find a negative cluster mass dependence of mass segregation effect (Figure 4).
    \item Mass segregation is visible in low-mass clusters simply as a result of dynamical friction time being shorter for more massive galaxies (Figure 5).
    \item Massive clusters may not show mass segregation mainly because the more centrally located galaxies tend to fall in their clusters relatively earlier and lose more mass during their longer orbital motion. In addition, because of their early arrival, their masses were not particularly large at the time of infall (Figures 5 and 6). 
\end{itemize}

Dynamical friction works as a drag force to segregate mass in clusters. It is natural to expect this to result in stellar mass segregation. However, tidal stripping acts as an obstacle and disturbs the mass segregation trend. The galaxies in the inner volume of the more massive clusters seem to have early infall time than those in the less massive clusters. The ratio of the effect of dynamical friction and mass stripping determines the visibility of mass segregation, and the trend is clearly related to the halo mass. 
Previous literature in general may have found no mass segregation trend either because they used a high mass cut or a higher range of redshift. 
These explanations can account for the discrepancies in the literature concerning the existence of mass segregation trends.

We cannot state that stellar mass segregation exists without knowing the specific conditions in the clusters. What we can conclude is that mass segregation does exist when certain conditions are met in the clusters and satellites. In this context, we conclude that the discrepancies found in the previous literature are not really disagreements after all; they all can be integrated into a single mass segregation scenario.\\

\section*{Acknowledgement}
S.K.Y. acknowledges support from the Korean National Research Foundation (NRF-2020R1A2C3003769). 
E. Contini acknowledges the National Key Research and Development Program of China (No. 2017YFA0402703), and the National Natural Science Foundation of China (Key Project No. 11733002). The supercomputing time for numerical simulation was kindly provided by KISTI (KSC-2014-G2- 003), and large data transfer was supported by KREONET, which is managed and operated by KISTI. Parts of this research were conducted by the Australian Research Council Centre of Excellence for All Sky Astrophysics in 3 Dimensions (ASTRO 3D) through project number CE170100013.

\bibliography{main}{}
\bibliographystyle{aasjournal}

\end{document}